\documentstyle[prd,aps,floats]{revtex}
\begin{document}
\draft

\input epsf \renewcommand{\topfraction}{0.8} 
\twocolumn[\hsize\textwidth\columnwidth\hsize\csname 
@twocolumnfalse\endcsname

\title{Post-inflationary brane cosmology} 
\author{Anupam Mazumdar}
\address{Astrophysics Group, Blackett Laboratory, Imperial College London,
 SW7 2BZ,~U.~K.}
\date{\today} 
\maketitle
\begin{abstract}
The brane cosmology has invoked new challenges to the usual Big Bang cosmology.
In this paper we present a brief account on thermal history of the post-inflationary
brane cosmology. We have realized that it is not obvious that the post-inflationary
brane cosmology would always deviate from the standard Big Bang cosmology. However, if it 
deviates some stringent conditions on the brane tension are to be satisfied. In this regard we 
study various implications on gravitino production and its abundance. We discuss 
Affleck-Dine mechanism for baryogenesis and make some comments on moduli and dilaton
problems in this context. 
\end{abstract}


\vskip2pc]

\section{Introduction}

Recently, there has been a great deal of interest in conceiving our universe to be a brane-world
embedded in a higher dimensional space-time \cite{rub}. Such a claim has a motivation from 
strongly coupled string theories \cite{horava,antoniadis}. The field theory limit of 
strongly coupled $E_{8}\times E_{8}$ hetrotic string theory (or M-theory) is believed to be an 
$11$ dimensional supergravity theory \cite{horava}. The $11$ dimensional world comprises 
two $10$ dimensional hypersurfaces
embedded on an orbifold fixed points. The fields are assumed to be confined to these hypersurfaces,
known as $9$ branes, which can be seen as forming the boundaries of the space-time. It has also
been shown that after compactifying the $11$ dimensional field theory on a Calabi-Yau three
fold it is possible to obtain an effective $5$ dimensional theory \cite{lukas}. The five 
dimensional space has a structure ${\cal M}_{5}={\cal M}_{4}\times S^{1}/Z_{2}$ and contains 
two three dimensional branes situated on the orbifold fixed points. The structure allows 
${\cal N}=1$ gauged supergravity in the bulk and on the orbifold fixed planes which could be a
realistic testable model for a particle physics phenomenology.

One the other hand there has been some interesting proposals to solve the hierarchy between the
two apparent scales the Planck and the electro-weak by introducing extra 
dimensions and also recognizing the higher dimensional Planck mass to be the fundamental
scale. This set-up does not require the world to be supersymmetric. However, this requires some
stringent conditions upon gravity residing in the brane and the bulk while the standard model 
fields are stuck to the observable world \cite{nima}. Then there came another twist in 
constructing models with extra dimensions. It has been shown that gravity can also be 
localized \cite{lisa}. The authors have demonstrated that in a background of a special 
non-factorizable geometry an exponential 
warp factor appears to the Poincar\'e invariant 3+1 dimensions.
The model consists of two $3$ branes situated rigidly along the 5th 
dimension compactified on a $S^{1}/Z_2$ orbifold symmetry. The space-time in the bulk is 
5 dimensional anti de-Sitter space and the two branes have opposite brane tensions.

All these new ideas invoke a great concern to the
cosmological evolution of the Universe and in this regard several authors have studied 
this question with a great emphasis on inflationary cosmology. It has also
been noticed that the two hypersurfaces with opposite brane tensions behave differently and 
they have different cosmology all together \cite{jing}. The most startling result was 
noticed to be departure 
from the usual four dimensional evolution equation on the brane \cite{binetruy,jim}. 
The presence of branes and 
the requirement that the fields are localized to the respective branes lead to a 
non-conventional brane cosmology, which requires lot more study. In this paper we concentrate 
upon one of the most important aspects of post-inflationary cosmology in our brane. 

The idea of cosmological Inflation has many virtues, solving a range of otherwise troubling 
problems. However, inflation leads to extremely cold Universe with entropy 
not sufficient to synthesize light elements. Hence, the Universe
has to be reheated upto a temperature sufficient enough to have nucleosynthesis, which is
close to ${\cal O}(\rm MeV)$. This is attained via the decay of coherent oscillations of 
the scalar field whose potential energy dominated the Universe before their decay. The Universe
ultimately reaches a thermal equilibrium and also radiation dominated era from where on it 
follows the standard picture of Big Bang cosmology. The maximum temperature attained 
during the radiation dominated era is known to be the reheat temperature and to match the bound 
coming from nucleosynthesis it should be at least more than 
${\cal O}(\rm MeV)$. The reheat temperature plays an important role in the standard Big Bang
cosmology and depending on the efficiency of reheating, the reheat temperature could be higher
or lower, and in either scenarios there could be interesting cosmological impacts. Interesting
results have been obtained for a Universe with a low reheat temperature, see Ref.~\cite{kolb}.

In this paper we will be estimating the reheat temperature
and then discussing various implications on our brane cosmology. Strictly speaking  
we will be treating the observable brane as a hypersurface. We will be assuming that at least in 
our brane supersymmetry is required to solve the hierarchy problem. In this regard we are 
closer to the string scenario where the effective four dimensional Lagrangian accommodates
${\cal N}=1$ supergravity with chiral and gauge multiplets. However, to keep the discussion 
quite general we will not a priori fix the volume of the six dimensional manifold or the length 
of the eleventh dimensional segment. This will give us ample choice to the five dimensional
Planck mass, which we denote here by $M_{5}$. We begin with a short introduction to  
chaotic inflation. We will then discuss reheat temperature, gravitino abundance, 
and towards the end we will briefly discuss the viability of Affleck-Dine mechanism for baryogenesis.
We then conclude our paper with some discussions. 

\section{chaotic inflation on the brane}
 
It has been noticed in Refs.~\cite{lukas1,binetruy} in the context 
of extra dimensions and the brane-world scenario that the effective four dimensional 
cosmology is non-trivial and could possibly deviate from a simple Big Bang cosmology.
Such a claim has a motivation from string theory which perhaps could lead us completely
different physics to the early Universe. It is thus interesting to study the consequences 
of string motivated early cosmology. However, due to advancement of the observational 
cosmology, it is no longer believed that the early cosmology 
is not well constrained. In this regard inflation which is still one of the best paradigms
of the early Universe is also constrained by the COBE data which has measured the temperature  
of the cosmic microwave background radiation and observed a small inhomogeneity which is 
one part in $10^{5}$. Thus it would be interesting to describe inflation in four dimensions
in the context of brane-world scenario. However, the noted deviation for the Friedmann
equation at very low temperatures leads to interesting consequences which we will be discussing 
in this section. Followed by the discussions in Refs. \cite{binetruy,jim} we notice that
the presence of an extra dimension, $y$, compactified on an orbifold $y=-y$ leads to 
extra terms in the Friedmann equation in the observable brane. We mention here only the 
leading terms
\begin{eqnarray}
\label{main}
H^2=\frac{8\pi}{3M_{\rm p}^2}\rho\left[1+\frac{\rho}{2\lambda}\right]+\frac{\Lambda_4}{3}\,,
\end{eqnarray}
where $\Lambda_4$ is a four dimensional cosmological constant, $\rho$ is the energy 
density of the matter stuck to the brane. In our discussion we will be assuming the four dimensional
cosmological constant to be precisely zero from the onset of inflation. The brane tension
$\lambda$ relates the four dimensional Planck mass to the five dimensional Planck mass via
\begin{eqnarray}
\label{main1}
M_{\rm p}=\sqrt{\frac{3}{4\pi}}\left(\frac{M_{5}^2}{\sqrt{\lambda}}\right)M_{5}\,.
\end{eqnarray}
It is evident that Eq.~(\ref{main}) leads to the usual relation $H=\sqrt{8\pi \rho/3M_{\rm p}^2}$
\footnote{We will frequently imply Eq.~(\ref{main}) to be a consequence of non-conventional
brane cosmology compared to the standard cosmology where $H=\sqrt{8\pi \rho/3M_{\rm p}^2}$.}
when $\rho < 2\lambda$. During nucleosynthesis in our brane this is precisely the criteria 
to be followed because at that time the expansion rate is 
determined by the energy density linear in $\rho$ in Eq.~(\ref{main}). This leads to 
constraining the brane tension $\lambda >(1{\rm MeV})^4$, and also the five dimensional
Planck mass $10^{4}{\rm GeV} < M_{5}$. However, as we shall soon notice that the 
upper bound on the five dimensional Planck mass will also be fixed by demanding that the
inflaton field would not take a value more than the Planck scale in four dimensions during
inflation. Here we briefly discuss some of the perspectives of the massive 
inflaton field with a potential $V=(m_{\phi}^2 \phi^2/2)$. It is important to mention that 
the presence of an extra dimension does not alter the conservation equation for the 
matter field stuck to the observable brane. 
\begin{eqnarray}
\label{conserve}
\dot \rho + 3 H (\rho+p)=0\,.
\end{eqnarray}
This is not always true, especially if there were a scalar field non-minimally coupled to gravity
living in the brane and the bulk, then the local conservation equation for a matter 
field stuck to the brane would not hold. Interesting cosmology will be discussed elsewhere,
but from now on we follow Eq.~(\ref{conserve}). This has an obvious consequence to the
scalar field dominating the early Universe during its potential dominated phase. The 
dominance of $\rho^2$ term in Eq.~(\ref{main}) leads to enhancing the Hubble friction term in 
Eq.~(\ref{conserve}). This naturally assists inflation provided a scalar field
is slowly evolving on a potential. It has been demonstrated in various 
Refs.~\cite{maartens,mendes,copeland} that chaotic inflation is possible in this non-conventional
scenario. However, to satisfy the COBE data on the observed density perturbations 
it is pertinent to constrain the mass of the inflaton field which is given by \cite{maartens}
\begin{eqnarray}
\label{con}
m_{\phi} \approx  5 \times 10^{-5}M_{5} \equiv 4 \times 10^{-5}M_{\rm p}^{1/3} \lambda ^{1/6}\,,
\end{eqnarray}
and the inflaton field \cite{maartens}
\begin{eqnarray}
\label{con1}
\phi_{\rm {cobe}}\approx 3 \times 10^{2}M_{5} \equiv 2 \times 10^{2}
M_{\rm p}^{1/3}\lambda^{1/6}\,,
\end{eqnarray}
where $\phi_{\rm {cobe}}$ is determined by the number of e-foldings required for 
generating an adequate density perturbations. While deriving the last equation the
authors in Ref.~\cite{maartens} have taken $N_{\rm cobe} \approx 55$ e-foldings, and,
we remind ourselves that the above bounds have been obtained while assuming that
the Hubble parameter is dominated by $\rho^2$ term in Eq.~(\ref{main}). 
If we do not want to plague the inflaton potential by 
non-renormalizable quantum corrections, we would require to begin inflation at a scale 
below the four dimensional Planck mass, which then allows $M_{5} < 10^{17}$ GeV, 
thus constraining the value of the five dimensional Planck mass to be 
$ 10^{17}{\rm GeV} < M_{5} < 10^{4}{\rm GeV}$, 
and similarly constraining the brane tension $10^{64}({\rm GeV})^4 < \lambda < 1({\rm MeV})^4$.

In this regard we notice that if we assume that our brane has ${\cal N}=1$ supersymmetry and
it is broken at a suitable scale in a hidden sector which is mediated to the observable
sector only through gravitational interactions such that it generates soft supersymmetry breaking
masses to the gravitino, sfermions and the gauginos, then to solve the gauge hierarchy problem
the gravitino mass should be around ${\cal O}(\rm TeV)$. If we also believe that 
supersymmetry in our world is a consequence of low energy string theory beyond 
supersymmetric standard model, then we get dilaton and moduli supermultiplets 
which will acquire a small mass $\sim {\cal O}(m_{3/2})$. It has been noted that their 
cosmology is very similar to the gravitinos \cite{nano,carlos}. Especially, the gravitino abundance 
is strongly connected to the reheat temperature, because they can be created after inflation. We will
be studying them in the next section.

\section{Reheat temperature}

One of the virtues of inflation is that the inflaton after the end of inflation becomes 
extremely homogeneous and begins coherent oscillations around the bottom of the potential.
For a massive inflaton field the average pressure vanishes during the oscillations and the 
energy density of the inflaton follows $\rho_{\phi} \propto a^{-3}$, where $a$ is the scale
factor. If the energy density of the decaying inflaton is larger than the brane tension, then
the Hubble expansion as a function of the scale factor can be written as 
\begin{eqnarray}
\label{hub}
H^2(a) \approx \frac{8\pi}{3M_{\rm p}^2}\frac{\rho_{\phi {\rm i}}^2}{2\lambda}
\left(\frac{a_{\rm i}}{a}\right)^6\,,
\end{eqnarray}
where we have denoted $\rho_{\phi \rm i}$ and $a_{\rm i}$ as the inflaton energy density 
and the scale factor at the beginning of the coherent oscillations. Depending on the 
decay rate of the inflaton the reheating process could be efficient or inefficient. The
only observed fact is that the reheat temperature should be more than $\sim {\cal O}({\rm MeV})$
to pave a successful nucleosynthesis. Equating Eq.~(\ref{hub}) to the decay rate $\Gamma_{\phi}$,
and then equating $\rho_{\phi}$ to the energy density of the 
relativistic species $\rho_{\rm r}=(\pi^2/30)g_{*}T_{\rm rh}^4$, where $g_{*}$ is the relativistic
degrees of freedom, we obtain the reheat temperature of the Universe \cite{anu}
\begin{eqnarray}
\label{rh}
T_{\rm rh} \approx \left(\frac{\Gamma_{\phi}M_{\rm p} \sqrt{\lambda}}{g_*}\right)^{1/4}
\approx \left(\frac{\Gamma_{\phi}M_{5}^3}{g_*}\right)^{1/4}\,.
\end{eqnarray}
For renormalizable couplings to the inflaton, the decay rate of the inflaton is estimated to be
$\Gamma_{\phi} \sim \alpha_{\phi} m_{\phi}$, where $\alpha_{\phi}$ is a dimensionless 
Yukawa coupling. The estimation of reheat temperature is given by \cite{anu}
\begin{eqnarray}
\label{rh1}
T_{{\rm rh}1} &\approx & 10^{-5/4}M_{\rm p}^{1/3}\lambda^{1/6}
\left(\frac{\alpha_{\phi}}{g_*}\right)^{1/4}\,,
\nonumber \\
&\approx & 10^{-5/4}M_{5}\left(\frac{\alpha_{\phi}}{g_*}\right)^{1/4}\,.
\end{eqnarray}
The inflaton field could also decay via gravitational interactions such as in the case 
of supergravity the decay rate is given by $\Gamma_{\phi} \sim m_{\phi}^3/M_{\rm p}^2$.
Then the estimation of reheat temperature is given by
\begin{eqnarray}
\label{rh2}
T_{{\rm rh}2} \approx 10^{-3}\left(\frac{\lambda}{g_{*}}\right)^{1/4} \approx 
10^{-3}\frac{M_{5}^{3/2}}{g_{*}^{1/4}M_{\rm p}^{1/2}}\,.
\end{eqnarray}
Thus in principle one can find a range of reheat temperatures for these two cases we have 
discussed, but before analysing them it is crucial to ensure whether the energy density of 
the thermal bath is more than the brane tension, otherwise our naive assumption behind
estimating the reheat temperature, namely Eqs.~(\ref{rh1}) and (\ref{rh2}) would be of no use. 
So, let us analyse the second scenario when the inflaton was decaying
via gravitational interactions. It is not difficult to realise that 
$\rho_{\rm r}(T_{{\rm rh}2}) \approx 10^{-11}\lambda$. This tells us that we were 
wrong behind our naive assumption. The conclusion is very simple and it suggests that if 
the inflaton is decaying via gravitational interactions, whatsoever be the value of the brane 
tension, after inflation we would always be in a regime where $\rho_{\rm r} < \lambda$, and,
thus we can safely assume the standard Big Bang lore 
$H \propto \sqrt{\rho}/M_{\rm p}$ instead of a non-conventional brane cosmology. This
has an interesting implication that if we were to decide a swift transition from non-conventional
inflationary cosmology to the standard Big Bang cosmology, then perhaps this could be easily
achieved via slow decaying of the inflaton field.

Now let us similarly analyse the first scenario when the inflaton field was decaying via 
Yukawa couplings. Thus to ensure that $\rho_{\rm rh}$ is greater than $\lambda$, we get
the following inequality 
\begin{eqnarray}
\rho_{\rm r}(T_{{\rm rh}1}) \equiv \frac{\alpha_{\phi}}{3}10^{-5} M_{5}^4 > \lambda 
\equiv \frac{3}{4\pi}\frac{M_{5}^6}{M_{\rm p}^2}\,,
\end{eqnarray} 
which will be true when the five dimensional Planck mass is constrained by
\begin{eqnarray}
\label{cond}
M_{5} < \frac{\sqrt{4\pi \alpha_{\phi}}}{10^{3}}M_{\rm p}\,,
\end{eqnarray}
which can be amply satisfied if we lower the five dimensional Planck mass. It is noticeable 
that the upper bound on the five dimensional Planck scale is at least an order of magnitude 
lower than the bound obtaining from beginning inflation below four dimensional Planck scale.  
This tells us an important message that if Eq.~(\ref{cond}) is satisfied, then we may safely 
assume Eq.~(\ref{hub}),  provided that the inflaton field is mainly decaying via the Yukawa 
couplings rather than pure gravitational couplings. This is an important conclusion which we 
have to bear in mind. From here onwards we will only concentrate upon inflaton decaying via 
Yukawa couplings which are not Planck mass suppressed. Only while discussing Affleck-Dine 
baryogenesis in non-conventional cosmology we will be assuming that the inflaton decaying via 
gravitational coupling.

Since we know that the transition from non-conventional cosmology to the standard cosmology 
should take place before nucleosynthesis, we must then estimate  the radiation temperature 
when the transition between non-conventional to the standard Big Bang cosmology takes place.
If we naively assume that the transition is instantaneous, then the transition temperature is
given by
\begin{eqnarray}
\label{trans}
T_{\rm transit} \approx \frac{M_{5}^{3/2}}{M_{\rm p}^{1/2}}\,,
\end{eqnarray}
where we have assumed $g_{*} \sim 300$. \footnote{For illustrating purposes we choose
$\alpha_{\phi} \sim 0.1$ and $g_{*} \sim 300$ at $T=T_{\rm rh}$. However, this
would also hold as long as the temperature is greater than the masses of the superpartners.
For temperatures below ${\cal O}(\rm MeV)$, $g_{*} \sim 3.36$.} 
It is evident from 
Eq.~(\ref{trans}) that 
higher the five dimensional Planck mass $M_{5}$ is, swifter the transition is.
For instance if $M_{5} \sim 10^{16}$ GeV, the transition from non-conventional to conventional
cosmology takes place when $T_{\rm transit} \approx 10^{14}$ GeV. It is also important to notice 
that this happens very close to the reheat temperature. However, for a low five dimensional 
Planck mass such as $M_{5}\sim 10^{6}$ GeV, the transition temperature could be as low as 
$T_{\rm transit} \leq 1$ GeV. Notice that this happens after the reheat temperature 
$T_{\rm rh} \approx 10^{4}$ GeV.

Now we can estimate the range of reheat temperature but only for a case when the
inflaton is decaying via Yukawa couplings.
\begin{eqnarray}
\label{bound}
10^{16}\left(\frac{\alpha_{\phi}}{g_{*}}\right)^{1/4}{\rm GeV}\geq T_{\rm rh} \geq 10^{3}\left(
\frac{\alpha_{\phi}}{g_{*}}\right)^{1/4} {\rm GeV}\,.
\end{eqnarray}
We notice that depending on the brane tension, or, the Planck mass in 
five dimensions, we can have completely different scenarios in the energy scale beyond 
nucleosynthesis. In this paper we will be discussing some of the concerning issues later on.
So far we have been assuming that the reheating would be prompt, or, at least the inflaton
would decay via the Yukawa couplings, or, the gauge couplings. However, if the reheating was 
not very prompt then one could imagine that during the process of reheating the Universe
could have a different temperature and perhaps more than the reheat temperature of the Universe.
In fact the suspicion is correct, and indeed the Universe can achieve a temperature higher
than the reheat temperature. This is due to the fact that if reheating was not 
prompt, then there could be a short spell where apart from the decaying inflaton, there 
could be a radiation content and also the decay products of the inflaton. The equation of state
of the matter content would neither mimic radiation nor a pressureless fluid, rather
completely different. It has been noticed that during this time in the standard cosmology
the temperature follows $T \propto a^{-3/8}$ \cite{chung}. Naively, one would expect similar 
kind of behaviour in the non-conventional case as well. However, it is not very difficult to 
realize that following the arguments in Ref.~\cite{chung}, the maximum temperature achieved
in non-conventional cosmology is $T_{\rm MAX} \approx {\cal O}(1)T_{\rm rh}$. It can 
also be noticed that the abundance of a stable massive particles are different in this scenario and
work in this direction is in progress. In the next section we describe the gravitino abundance in
non-conventional cosmology.

 
\section{Gravitino production and their abundance}

If we imagine that supersymmetry plays an important role in the early Universe then our scope of 
discussion enhances a lot, and, out of which the gravitino production and their abundance 
gets most of our attention. In the gravity mediated supersymmetry breaking the gravitino
gets a mass around ${\cal O}(\rm TeV)$. Since their couplings to other particles are Planck mass
suppressed, the life time of gravitino at rest is quite long $\tau_{3/2} \sim M_{\rm p}^2/m_{3/2}^3
\sim 10^{5}(m_{3/2}/TeV)^{-3}$ sec \cite{cline}. We know that successful nucleosynthesis
depends on the ratio of the number density of baryons to photons. The gravitino decay products 
could easily change this ratio. Their decay products such as gauge bosons and its 
gaugino partners, or, high energy photons could generate a large entropy which would 
heat up the photons compared to $\tau$ and $\mu$ neutrinos. The abundance of neutrinos
essentially determines the $^{4}{\rm He}$ abundance, and this way even if the gravitinos 
are not the lightest supersymmetric particles they could cause considerable harm. 
It was first pointed out in Ref.~\cite{weinberg} that the 
gravitino mass should be larger than $\sim 10$ TeV in order to keep the successes of 
the Big Bang nucleosynthesis. On contrary if the gravitinos were 
stable and if their mass exceeded $1$ KeV, they could easily overclose the Universe in 
absence of inflation \cite{pagel}.   
Thus studying gravitino abundance is a paramount in this regard. Especially if 
their energy density is more than the brane tension, then some interesting
physics may take place which were not present in the standard cosmology. 
With this hope we explore the gravitino abundance in this section. 

This section will be reviewed from the results obtained in Ref.~\cite{anu}. The gravitinos 
can be created in a thermal bath. For  helicity $\pm 3/2$ gravitinos the cross-section 
is given by $\sigma \propto (g^2/M_{\rm p}^2)$, where $g$ is the gauge coupling constant
and for helicity $\pm 1/2$ gravitinos the cross-section is given by 
$\sigma \propto (g^2m_{g}^2/M_{\rm p}^2m_{3/2}^2)$ \cite{fayet}. In either
case we notice that the cross-section is suppressed by the four dimensional Planck mass. However,
in the latter case the cross-section could be  actually suppressed by the supersymmetric breaking
mass scale $M_{\rm s}^2$ which determines the mass of the gravitino $m_{3/2} \sim 
M_{\rm s}^2/M_{\rm p}$, provided $m_{\rm g} \neq m_{3/2}$.
In order to study the gravitino abundance we need to study the Boltzmann equation
for the gravitino number density \cite{kolb1}
\begin{eqnarray}
\label{imp0}
\frac{dn_{3/2}}{dt}+3Hn_{3/2}=\langle\Sigma_{\rm tot} v_{\rm rel}\rangle n^2_{\rm rad}
-\frac{m_{3/2}}{\langle E_{3/2}\rangle}\frac{n_{3/2}}{\tau_{3/2}}\,,
\end{eqnarray}
where $\langle ...\rangle$ represents thermal average, $n_{\rm rad}$ is the number 
density of relativistic particles $n_{\rm rad}\propto T^3$, $v_{\rm rel}$ is the 
relative velocity of the scattering radiation which in our case $\langle v_{\rm rel}\rangle =1$,
and the factor $m_{3/2}/\langle E_{3/2}\rangle$ is the average Lorenz factor. We notice
that in the radiation era non-conventional brane cosmology gives the following Hubble 
expansion
\begin{eqnarray}
\label{imp1}
H \approx \left(\frac{4\pi^5}{3}\right)^{1/2}\frac{g_{*}}{30}\frac{T^4}{\sqrt{\lambda}M_{\rm p}}\,.
\end{eqnarray} 
In supersymmetric version $g_{*} \sim 300$ if the reheat temperature is more than the masses 
of the superpartners. It is worth mentioning that the scale factor during the radiation era
follows $ a(t) \propto t^{1/4}$, which is contrary to the standard Big Bang scenario where 
$a(t) \propto t^{1/2}$. However, we must not forget that the derivation is based on the fact 
that we are in a regime where $\rho > 2\lambda$. In Eq.~(\ref{imp0}), after the 
end of inflation the first term in the right-hand side dominates the second. If we assume 
the adiabatic expansion of the Universe $a\propto T^{-1}$, then we can 
rewrite Eq.~(\ref{imp0}) as $Y_{3/2}=(n_{3/2}/n_{\rm rad})$.
\begin{eqnarray}
\label{abu}
\frac{dY_{3/2}}{dT} \approx -\frac{\langle \Sigma_{\rm tot}\rangle n_{\rm rad}}{HT}\,.
\end{eqnarray}
We notice that we can integrate the temperature dependence from this equation and
the expression is almost independent of the temperature.
we mention here that the above expression is exactly the same as in the standard 
Big Bang case \cite{kolb1}. However,  Eq.~(\ref{abu}) does not produce the correct 
value for $Y_{3/2}$, since the true conserved quantity is the entropy per comoving 
volume. In our case if we assume the gravitinos do not decay within the time frame 
we are interested in, then we may be able to get an expression for the gravitino abundance 
at two different temperatures 
\begin{eqnarray}
Y_{3/2}(T) \approx \frac{g_{*}(T)}{g_{*}(T_{\rm rh})}\frac{n_{\rm rad}(T_{\rm rh})
\langle \Sigma_{\rm tot}\rangle}{H(T_{\rm rh})}\,.
\end{eqnarray}
Here we  assume that the initial abundance of gravitinos at $T_{\rm rh}$ is known to us, and
the dilution factor $g_{*}(T)/g_{*}(T_{\rm rh})$ takes care of the decrease in the
relativistic degrees of freedom. For a rough estimate we assume the total cross-section 
$\Sigma_{\rm tot} \propto 1/M_{\rm p}^2$, and, $n_{\rm rad}(T_{\rm rh}) \propto T_{\rm rh}^3$, 
we finally get an expression for the gravitino abundance at temperature $T$ \cite{anu}
\begin{eqnarray}
\label{life}
Y_{3/2}(T<{\rm MeV}) \approx 10^{-3} \frac{\sqrt{\lambda}}{T_{\rm rh} M_{\rm p}}\,,
\end{eqnarray}
where we have assumed the dilution factor to be $\sim 10^{-2}$.
If we assume that the inflaton decays via the Yukawa couplings, then with the help of
Eq.~(\ref{rh1}) we get a simple expression
\begin{eqnarray}
\label{life1}
Y_{3/2}(T) &\approx & 10^{-7/4}\frac{\lambda^{1/3}}{M_{\rm p}^{4/3}}
\left(\frac{g_{*}}{\alpha_{\phi}}\right)^{1/4}\,, \nonumber \\
&\approx &10^{-7/4}\left(\frac{M_{5}}{M_{\rm p}}\right)^2
\left(\frac{g_{*}}{\alpha_{\phi}}\right)^{1/4}\,,
\end{eqnarray}
where $g_{*}$ is actually evaluated at $T_{\rm rh}$, and, the abundance expression is true for 
temperatures below MeV.
The above expression for the abundance of the gravitinos is quite important. It tells us 
directly that higher the value of $M_{5}$ is, higher the abundance is. For an example if 
$M_{5} \sim 10^{15}$ GeV, the abundance is roughly $Y_{3/2}\approx 10^{-8}$, which is 
more than the required acceptable bound $Y_{3/2} \leq 10^{-10}$ \cite{sarkar} for a 
successful nucleosynthesis. This puts severe constraint on the five dimensional Planck mass
\begin{eqnarray}
\label{bound1}
M_{5} \leq 10^{-5}M_{\rm p}\,,
\end{eqnarray} 
where we have taken $\alpha \sim 10^{-1}$, $g_{*} \sim 300$,
and the brane tension is given by
\begin{eqnarray}
\label{bound2}
\lambda \leq 10^{-31}M_{\rm p}^4\,.
\end{eqnarray}
Thus we see that introducing supersymmetry in the brane leads to lowering  the upper bound on 
the five dimensional Planck mass and also the brane tension in the brane-world set-up.

Following Eq.~(\ref{imp0}) we notice that as the Universe expands the Hubble time 
$H^{-1} \rightarrow \tau_{3/2}$, and at the time of decay the last term starts dominating 
the rest. Setting $M_{3/2}/\langle E_{3/2}\rangle =1$, it is then easy to estimate the 
gravitino abundance, which yields
\begin{eqnarray}
\label{abun}
Y_{3/2}(t) \approx 10^{-3}\frac{\sqrt{\lambda}}{T_{\rm rh}M_{\rm p}}e^{-t/\tau_{3/2}}\,,
\end{eqnarray}
where $\tau_{3/2}$ is the life time of the gravitino. We notice that as the five dimensional
Planck mass $M_{5}$ increases, the transition to the conventional cosmology becomes 
swifter. This suggests that the gravitinos with mass $\sim {\rm TeV}$ would 
decay close to $T_{\rm decay} \approx {\cal O}({\rm MeV})$. This is because the Universe
follows the conventional cosmology with $t \sim (T/{\rm MeV})^{-2}{\rm sec}$ after the transition. 
On the other hand if $M_{5}$ is smaller then one could expect that the temperature at which 
the gravitinos decay would be higher, determined by Eq.~(\ref{imp1}). However, one could also confirm
that throughout the decay life time the evolution of the Universe would not be the same as 
determined by Eq.~(\ref{imp1}), because of the transition from $a\propto t^{1/4}$ to 
$a\propto t^{1/2}$ taking place before the gravitinos could decay. If this is so then it is possible to 
roughly estimate the temperature at which the gravitinos would decay denoted by $T_{\rm decay}$.
\begin{eqnarray}
\label{dodgy}
\frac{\tau_{3/2}}{\tau_{\rm rh}} \approx \frac{T_{\rm rh}^4}{T_{\rm decay}^2 T_{\rm transit}^2}\,,
\end{eqnarray}
where $\tau_{\rm rh}$ can be assumed to be the time when the gravitinos are created in a 
thermal bath, which can be estimated from Eq.~(\ref{imp1}). With the help of Eqs.~(\ref{rh1}),
(\ref{trans}) and (\ref{imp1}), it can be shown that the temperature at which the gravitinos decay
is almost insensitive to the five dimensional Planck mass
\begin{eqnarray}
T_{\rm decay}&\approx &10^{-25/2} \left(\frac{\tau_{3/2}}{\rm sec}\right)^{-1/2} 
\left(\frac{M_{\rm p}}{\rm GeV}\right)^{1/2} {\rm GeV}\,, \nonumber \\
&\approx &10^{-3}\left(\frac{\tau_{3/2}}{\rm sec}\right)^{-1/2} {\rm GeV}\,.
\end{eqnarray}
For a TeV mass gravitino, whose life time is around $\tau_{3/2}\sim 10^{4} {\rm sec}$, which leads to
the decay temperature $T_{\rm decay} \approx 0.01 {\rm MeV}$. Thus the gravitinos
decay very late and even if the five dimensional Planck mass is small they could be potentially 
threatening to nucleosynthesis. The only way to save nucleosynthesis is to have small gravitino
abundance, which is possible only if we lower $M_{5}$ considerably according to Eq.~(\ref{bound1}).
The gravitinos are important from several points of view. Their decay products would always
contain the lightest supersymmetric particles and they could still survive in the form of 
cold dark matter. Usually while the gravitinos decay they generate an entropy as discussed 
earlier and they could as well wash away previously obtained baryon asymmetry in the Universe. 
The gravitinos decaying via CP non-conserving interactions as discussed in Ref.~\cite{jcline} 
could be an interesting scenario as to regenerate baryogenesis.

\subsection{Non-perturbative aspects of the gravitino production} 

So far we have been assuming that the gravitinos have been created in a thermal bath at 
a temperature close to the reheat temperature. However, it has been very recently realized 
that the gravitinos like other particles could as well be created non-perturbatively \cite{maroto}.
Though the authors concentrated only upon the helicity $\pm 3/2$ case, soon after that the
mechanism for exciting the other half $\pm 1/2$ were discussed in Ref.~\cite{kallosh}.
It was pointed out that helicity $\pm 1/2$ gravitinos were created more abundantly than the
$\pm 3/2$ case, because helicity $\pm 1/2$ gravitinos essentially eat the Goldstino mass,
which for a single chiral field is nothing but the supersymmetric partner of the inflaton,
usually called as inflatino. The inflatino mass is not suppressed by the four dimensional 
Planck mass and as a result the time varying mass contributing to the Goldstino could 
boost the production. In the brane-world scenario we suspect that the non-perturbative production
of the gravitinos could be even more important.

To keep this discussion general we discuss some of the key features of preheating. The 
elaborate idea has been discussed in Refs.~\cite{linde}. After the end of inflation the 
scalar field begins oscillating around the bottom of the potential when mass of the 
inflaton is comparable to the Hubble expansion $m \sim H$. The energy density
of the inflaton field $\phi$ decreasing in a same way as a non-relativistic particle of mass
$m$, where $\rho_{\phi} =\dot \phi^2/2 + m^2 \phi^2 /2 \sim a^{-3}$, and $a$ is the scale factor of 
the Universe. During the homogeneous oscillations of the scalar field the Universe acts as a matter
dominated era where the scale factor grows like $a(t) \approx a_{\rm i}(t/t_{\rm i})^{2/3}$
in the standard cosmology. In non-conventional cosmology from Eq.~(\ref{hub}), it is 
clear that the scale factor would grow as $a(t) \approx a_{\rm i}(t/t_{\rm i})^{1/3}$. Thus in a
non-conventional cosmology the scale factor grows slowly compared to that in the standard
case, and this has lots of interesting implications. Here we briefly sketch some of them.
It is clear that the oscillations in $\phi$ are sinusoidal with a decreasing amplitude 
$\phi(t) \sim (1/m\sqrt{t})$ in the non-conventional scenario. However, in the standard cosmology
the amplitude of the oscillations follows $\phi(t) \sim (1/mt)$, thus the oscillations
die down faster in the conventional case compared to the non-conventional. For non-perturbative
creation of particles it is important to have a very high amplitude oscillations. It is worth
mentioning that one of the criterion for shutting off non-perturbative production is via 
decaying amplitude of the oscillations. 

In the discussion of non-perturbative creation of particles, there is a very useful time 
varying parameter, known as ``$q$'' parameter, given by $q=\alpha_{\phi}^2 \phi(t)^2/m^2$, 
where $\alpha_{\phi}$ is the Yukawa coupling. The production of either bosons and fermions 
depend very crucially upon this parameter. For the bosons the occupation number 
for a given momentum mode $k$ grows as $n_{k}(t) \propto e^{2\mu_{k}mt}$,
where $\mu_{k} =((q/2)^2-(2k/m-1)^2)^{1/2}$ is know as the Floquet index. In  
non-conventional case it is possible to have $q$ parameter bigger compared to that in 
the standard case. An another important factor is redshifting of the momentum, in
the case of non-conventional cosmology this effect is again weaker compared to that of
the conventional case due to slower expansion rate. A similar argument can also be given
for fermion creation. It has been noticed that the massive fermion production rate 
follows $\rho_{\rm fermion}\propto q$ \cite{peloso}. Thus clearly showing the importance of
$q$ parameter. 

The general argument behind production of the gravitinos is more or less the same as any other
fermions. However, it has been realized that the production of particles does depend on a specific
type of inflationary model. The discussion here primarily based on inflaton being massive and 
derived from a superpotential $W=\lambda \Phi^3$. In this case it has been noticed that the 
number density of the gravitino can be $n_{3/2} \sim H_{\rm i}^3$, where $H_{\rm i}$ is the Hubble 
parameter during inflation \cite{kallosh}. In non-conventional case $H_{\rm i}$ would be larger
than that of the conventional case and naively one would expect enhanced production of gravitinos. 
However, there exists other interesting 
superpotentials such as $W=\lambda \Phi (\chi^2-\chi_{0}^2)$, which leads to a potential 
similar to hybrid inflationary model \cite{hybrid}, where the production of gravitinos
has been noted to be $n_{3/2} \sim  (2\lambda \chi_{0})^3$, where $2\lambda \chi_0$ represents
the effective mass of the oscillating field \cite{mar}. One could as well find the abundance of
the gravitinos in this case, which is given by $n_{3/2}/s \sim (\lambda/\chi_0)T_{\rm rh}$ 
\cite{mar}. Thus the non-perturbative production does depend on the model parameters as well,
but whatsoever be the situation, for a given inflationary model if the production is larger
compared to that of the perturbative production, one would require to invoke the bounds on 
$M_{5}$ and $\lambda$. There is also an interesting proposal to create the gravitinos
after reheating \cite{lyth}, and it would be again interesting to investigate their abundance
in non-conventional cosmology.

\section{Viability of Affleck-Dine baryogenesis in non-conventional cosmology}

In this section we discuss briefly baryogenesis in the context of supersymmetry.
This discussion is mainly to illustrate that the physics of the early Universe could be
different in the brane-world cosmology. Our discussion will be mainly based on 
Refs.~\cite{ellis1,ellis2,anu},
but here we will make some additional comments. As we know that there are three main requirements 
for producing net baryon asymmetry in the Universe, baryon number violating interactions,
C and CP violations and a departure from thermal equilibrium \cite{shak}. Grand Unified Theories
(GUT) predict baryon number violation interaction at tree level and decay of a massive
Higgs bosons in out of equilibrium could give rise to the observed baryon asymmetry which is 
roughly one part in $10^{10}$. Similarly in the electro-weak baryogenesis it is possible that
nonperturbative effects could give rise to processes which preserve $B-L$, however, violates $B+L$,
where $B$ represents the baryon asymmetry and $L$ represents the lepton asymmetry. It is 
however possible to get baryon asymmetry from the lepton asymmetry \cite{fugi}. Thus there
exists many ways to extract baryon asymmetry and the expansion of the Universe fulfills the 
third criteria of out-of equilibrium decay. One could imagine this to happen during preheating
itself and one such example has been described in Ref.~\cite{peloso}. In this section we
would concentrate solely upon a supersymmetric mechanism for generating baryon asymmetry 
through the decay of sfermion condensate proposed in Ref.~\cite{ad}, known as 
Affleck-Dine (AD) mechanism. This mechanism depends crucially upon the total evolution of
the AD field starting from inflation till the era when supersymmetry breaking effects become 
important and in our case it could be an interesting example to study the difference between 
the standard and the non-conventional cosmology.

In the original AD scenario it was assumed that initially the sfermions would have to have 
large vacuum expectation values along the flat directions of the scalar potential.
However, the flatness could be spoiled if there were bout of inflation, because inflation
generically leads to a mass correction of the order of ${\cal O}(H^2)$ to the sfermions, this is
particularly true for $F$-type supersymmetric inflationary models. However, accidental cancellations
could occur in the inflaton sector due to a special choice of superpotential \cite{ewan}, which would
prevent effective mass gaining to the inflaton. The second potential threat comes from  
non-renormalizable terms in the superpotential, which would inevitably lift the flat directions.
The two points which we have mentioned here have their bad consequences to bringing down the
AD field very close to the minimum of the potential and as a result preventing the
AD baryogenesis. If ${\cal O}(H^2)$ correction to the mass of the AD field is negative by
a choice of Kahler potential \cite{lis}, then the AD field could sit at a minimum during
inflation which would be different from zero. In this section we would consider a simple toy 
model and estimate the initial value of the AD field.

After the end of inflation eventually the bare
mass term for the AD field $\sim {\cal O}(m_{3/2})$ dominates the four dimensional Planck 
mass suppressed non-renormalizable corrections and begins oscillating when $H \sim \tilde m
\equiv m_{3/2}$, where we have denoted $\tilde m$ as a mass of the AD field.
For our purpose we will simply assume $\rho_{\psi} \approx \tilde m ^2 \psi^2 $.
However, we must mention that our case is quite complicated. The reason is following.
An important condition to realize the AD baryogenesis is that the thermalization due to the
decay products of the inflaton field must take place after the decay of the AD field, and, 
$\rho_{{\rm r}\phi} > \rho_{\psi}$, where $\rho_{{\rm r}\phi}$ is the energy density in radiation
after the inflaton decay. This is a very stringent condition otherwise whatever
baryon asymmetry generated prior to thermalization would be washed away. 
To prevent this happening, the decay rate of the inflaton should be 
sufficiently small. This tells us that if the inflaton decays much earlier via the Yukawa
couplings then there is no way we can realize the AD baryogenesis. Thus, the inflaton 
must decay via the gravitational couplings, which is quite slow enough. However, as we 
have learnt in our earlier discussions that if the inflaton field decays via the gravitational
coupling then after thermalization the brane tension would dominate the energy density
of the theramlized plasma and the Universe would behave as if it were in the standard case
without any non-conventional term in the evolution equation. This means that if we begin with 
energy density in the inflaton field more than the brane tension, then while the inflaton
is oscillating there is a transition from non-conventional cosmology to the standard
cosmology. We assume that the transition takes place instantly and this happens when
$\rho_{\phi}\approx m_{\phi}^2\phi^2(a_{\phi}/a)^3 \sim \lambda$ at
\begin{eqnarray}
\label{trick1}
a =a_{\lambda} =\left(\frac{m_{\phi}^2\phi^2}{\lambda}\right)^{1/3}a_{\phi}\,,
\end{eqnarray}
where $a_{\phi}$ is the scale factor at the time when the inflaton begins oscillations,
$a$ is just the scale factor and $m_{\phi}$ denotes the mass of the inflaton. We picturize a 
situation where the Universe began with a 
non-conventional cosmology, then after the end of inflation the inflaton begins oscillating, 
but the Universe is still non-conventional. When the Hubble parameter drops to a 
value $H \sim \tilde m$ the oscillations in the AD field begins and at this time also the 
Universe is non-conventional.
However, soon after oscillations in the AD field is induced, the transition from  
non-conventional to the standard cosmology paves its way. Since the mass of the AD field 
is $\tilde m \sim m_{3/2} < m_{\phi}$
small compared to the mass of the inflaton, the oscillations in the AD field begin after 
the inflaton oscillations. This can be estimated by taking $H\sim \tilde m$. Since this happens
when the Universe is non-conventional; $H \approx (m_{\phi}^2\phi^2/M_{\rm p}\sqrt{\lambda})
(a_{\phi}/a)^3 \sim \tilde m$. We can estimate the scale factor when this happens
\begin{eqnarray}
\label{trick2}
a=a_{\psi} =\left(\frac{\sqrt{\lambda}}{M_{\rm p}\tilde m}\right)^{1/3}a_{\lambda}\,,
\end{eqnarray}
where we have used Eq.~(\ref{trick1}). It can be verified easily that $a_{\psi}<a_{\lambda}$.
However, this restricts the five dimensional Planck mass $M_{5}< 10^{14}$ GeV.
\footnote{For numerical estimations we have assumed $\tilde m \approx m_{3/2} \sim 1$ TeV.} After
the AD field begins oscillations we assume that the transition takes place at a given scale
factor by Eq.~(\ref{trick1}). After $a_{\lambda}$ the cosmology becomes the standard one and the Hubble 
rate is given by $H\propto \sqrt{\rho}/M_{\rm p}$. In our set-up the inflaton decays when the
Universe is already in the standard cosmology, thus we can estimate the scale factor when this happens 
by equating the Hubble parameter to the decay rate of the inflaton; 
$ H \approx (m_{\phi}\phi/M_{\rm p})(a_{\phi}/a_{\lambda})^{3/2} 
(a_{\lambda}/a)^{3/2} \sim \Gamma_{\phi}=(m_{\phi}^3/M_{\rm p}^2)$. Notice that the decay rate 
of the inflaton is via the gravitational coupling. This yields
\begin{eqnarray}
\label{trick3}
a=a_{{\rm d}{\phi}}=\left(\frac{\lambda M_{\rm p}^2}{m_{\phi}^6}\right)^{1/3}a_{\lambda}\,.
\end{eqnarray}
It can be verified that $a_{\phi}< a_{\psi} < a_{\lambda}< a_{{\rm d}\phi}$. 
During the oscillations of the AD field, the energy density decreases in the same fashion as in the  
case of inflaton. We can estimate the energy density in the AD field by
\begin{eqnarray}
\label{trick4}
\rho_{\psi}&=&\tilde m^2\psi_{0}^2\left(\frac{a_{\psi}}{a}\right)^3\, \nonumber \\
&=& \frac{\tilde m\sqrt{\lambda}\psi_{0}^2}{M_{\rm p}}\left(\frac{a_{\lambda}}{a}\right)^3\,,
\end{eqnarray}
where we have assumed Eq.~(\ref{trick2}). In order to decide which field decays first we have to 
compare the two decay rates. The decay rate of the AD field can be taken to be $\Gamma_{\psi}
\sim (\tilde m^3/\psi^2)$ \cite{ellis2}. The value of $\psi$ can be estimated from 
Eq.~(\ref{trick4}). Thus the condition that $\Gamma_{\phi}/\Gamma_{\psi} >1$ leads to 
\begin{eqnarray}
\left(\frac{a}{a_{\lambda}}\right)^3 <\frac{m_{\phi}^3M_{5}^3\psi_{0}^2}{M_{\rm p}^4 \tilde m^4}\,.
\end{eqnarray}
Following the values for $m_{\phi} \sim 10^{-5}M_{5}$, $M_{5}\sim 10^{13}$ GeV, and 
$\psi_{0} \sim 10^{4}M_{5}$, we get $a/a_{\lambda} < 10^{6}$. 
On the other hand for the same parameters we can also 
estimate $a_{{\rm d}\phi}/a_{\lambda} \sim 10^{4}$. Notice that the initial amplitude of the
AD field $\psi_{0}$ is taken larger than $\phi_{\rm COBE}$. This tells us that the 
inflaton will decay first. But an important factor is that  
thermalization due to the decay of the inflaton field must happen after the full decay of the AD
field.     

Once the Universe becomes radiation dominated, the energy density of the relativistic decay products
of the inflaton can be given by
\begin{eqnarray}
\label{bore0}
\rho_{{\rm r}\phi}&=&\frac{m_{\phi}^6}{M_{\rm p}^2}\left(\frac{a_{{\rm d}\phi}}{a_{\lambda}}\right)^4
\left(\frac{a_{\lambda}}{a}\right)^{4}\, \nonumber \\
&=&\left(\frac{\lambda^{4/3}M_{\rm p}^{2/3}}{m_{\phi}^2}\right)\left(\frac{a_{\lambda}}{a}\right)^4\,,
\end{eqnarray}
where we have used Eq.~(\ref{trick3}).
Notice that while deriving the above equation we have assumed the standard cosmology and also note that
during the radiation era $\rho \propto a^{-4}$. In the radiation dominated era the Hubble 
parameter in the standard cosmology is given by
\begin{eqnarray}
\label{bore1}
H=\left(\frac{\lambda^{2/3}}{m_{\phi}M_{\rm p}^{2/3}}\right)\left(\frac{a_{\lambda}}{a}\right)^{2}\,.
\end{eqnarray}
Now we must estimate when the AD field decays, following Refs.~\cite{ellis1} and \cite{ellis2} 
we equate $H\sim \Gamma_{\psi} \equiv \tilde m^3/\psi^2$.
This takes place when the scale factor is given by
\begin{eqnarray}
a = a_{{\rm d}\psi}=\left(\frac{\lambda^{7/6}\psi_{0}^2}{\tilde m^4 m_{\phi} M_{\rm p}^{5/3}}\right)^
{1/5}a_{\lambda}\,.
\end{eqnarray}
It can be verified that $\rho_{{\rm r}\phi}(a_{{\rm d}\psi}) > \rho_{\psi}(a_{{\rm d}\psi})$.
Now we have to make sure that the thermalization of the inflaton field happens after the decay of the
the AD field. For that we need to estimate the thermalization rate of the inflaton field.
Following the arguments given in Refs.\cite{ellis1} and \cite{ellis2} we get
\begin{eqnarray}
\Gamma_{\rm T}&\sim &n_{\phi}\sigma \sim m_{\phi}\phi^2\left(\frac{a_{\phi}}{a}\right)^3
\left(\frac{\alpha^2}{m_{\phi}^2}\right)\left(\frac{a}{a_{{\rm d}\phi}}\right)^2\, \nonumber \\
&\sim &\alpha^2\left(\frac{\lambda^{1/3}m_{\phi}}{M_{\rm p}^{4/3}}\right)
\left(\frac{a_{\lambda}}{a}\right)\,,
\end{eqnarray}
where $n_{\phi}$ is the number density of the relativistic particles, $\sigma$ is the cross-section
and $\alpha$ is the fine structure constant. The thermalization of the Universe occurs when
$\Gamma_{\rm T} \sim H$, where $H$ has already been estimated in Eq.~(\ref{bore1}).
\begin{eqnarray}
\label{cane}
a_{\rm T}=\alpha^{-2}\left(\frac{\lambda^{1/3}M_{\rm p}^{2/3}}{m_{\phi}^2}\right)a_{\lambda}\,.
\end{eqnarray}
At this point we can also check that $a_{{\rm d}\psi} < a_{\rm T}$ for $m_{\phi} \sim 10^{-5}M_{5}$, 
and $\alpha \sim 10^{-3/2}$. The condition is satisfied for any reasonable 
value of $\psi_0$ less than the four dimensional Planck mass.

At $a_{\rm T}$ we can compute the final baryon to entropy ratio given by \cite{ad}
\begin{eqnarray}
\label{ratio}
n_{\rm B}=\epsilon \left(\frac{\psi_{0}^2}{M_{\rm G}^2}\right)\frac{\rho_{\psi}}{\tilde m}\,,
\end{eqnarray}
where $\psi_0$ is the initial amplitude of the sfermion oscillations, $M_{\rm G}$ can be 
assumed to be an intermediate scale, could be supersymmetric grand unification scale and 
$\epsilon(\psi_{0}^2/M_{\rm G}^2)$ is the net baryon number generated by the decay of $\psi$.
At $a_{\rm T}$ the entropy density can be calculated with the help of Eqs.~(\ref{bore0}) and 
(\ref{cane}).
\begin{eqnarray}
s=\left(\rho_{{\rm r}\phi}(a_{\rm T})\right)^{3/4} \approx \frac{\alpha^{6}m_{\phi}^{9/2}}
{M_{\rm p}^{3/2}}\,,
\end{eqnarray}
and finally the baryon to entropy ratio can be given by 
\begin{eqnarray}
\label{final}
\frac{n_{\rm B}}{s}=\frac{\epsilon \psi_{0}^4 m_{\phi}^{3/2}}{M_{\rm G}^2\sqrt{\lambda}M_{\rm p}^{3/2}}
\equiv \frac{\epsilon \psi_{0}^4 m_{\phi}^{3/2}}{M_{\rm G}^2 M_{5}^3 M_{\rm p}^{1/2}}\,.
\end{eqnarray}
It is noticeable that the baryon to entropy ratio does not depend on $\tilde m$. However, it does 
depend on the brane tension and the initial amplitude of the AD field oscillations. The last step
in the above equation has been been expressed in terms of the five dimensional Planck mass.
For an example, we may take $M_{\rm G} \sim 10^{15}$ GeV, $m_{\phi}\sim 10^{-5}M_{5}$, we get
an estimation of the initial amplitude of oscillations in the AD field
\begin{eqnarray}
\psi_{0}=\left(\frac{10^{37}}{\epsilon}\right)^{1/4}
\left(\frac{M_{5}}{\rm GeV}\right)^{3/8}~{\rm GeV}\,,
\end{eqnarray}
where we have taken the observed baryon to entropy ratio to be $n_{\rm B}/s \sim 10^{-10}$.
It is evident that the value of $\psi_{0}$ is more than $\phi_{\rm COBE} \approx 10^{2}M_{5}$.
However, for smaller values of $M_{5}$ the amplitude could be comparable to $\phi_{\rm COBE}$.
In that case, situation could be different. Here we have implicitly assumed that the AD field
decays after the decay of the inflaton. For smaller values of $\psi_{0}$, the situation could be
reversed, in that case the AD field would decay before the inflaton decay. In such a case,
the entropy produced would be simply given by the inflaton decay and we do not have to bother about 
actual thermalization of the relativistic particles.

Here we would like to make remark upon the baryon to entropy ratio obtained in Ref.~\cite{anu}. 
The treatment was slightly different there and not fully correct, because the author did not consider 
the transition from non-conventional to the standard cosmology, which happens while the inflaton
is still oscillating. This is a crucial point which also makes it different from the AD mechanism 
in the standard cosmology.

\section{$M_{5}$ from M-theory vacua}

So far we have been setting the value of $M_{5}$ at our will without much justification.
However, the five dimensional Planck mass should be fixed from the observed four dimensional
Planck mass, the string scale and the size of the internal dimensions. These scales are  
also closely related to the strength of the gauge couplings. In this regard we will only
focus upon  M-theory on $S^{1}/Z_{2}$. As we have earlier discussed in the introduction 
that the field theory limit of the strongly coupled string theory (or M-theory) has been 
shown to be $11$ dimensional supergravity compactified on a manifold with 
boundaries expressed as $S^{1}/Z_{2}\times {\rm CY}$ \cite{horava,witten}, where $S^{1}/Z_{2}$ is
a line segment of size $\pi r$ and ${\rm CY}$ is a Calabi-Yau manifold with a volume $V$ in $6$ 
dimensions. In this context the five dimensional Planck mass is given by
\begin{eqnarray}
M_{5}=\frac{M_{11}^3}{V^{-1/3}}\,,
\end{eqnarray}
where $M_{11}$ is the string scale in $11$ dimensions. In fact there are three cases which 
have been discussed in several literatures. Various concerning  
phenomenological issues were already discussed in this regard in Ref.~\cite{benakli,benakli2}. 
If the string scale lies above $1$ TeV then one can imagine that the high energy theory is
supersymmetric and supersymmetry is broken in our brane at a scale around $1$ TeV. In this
regard our earlier discussions would hold true. We
summarise three cases here.

{\it Case $1 \rightarrow$}
$M_{11} \sim 10^{16}$ GeV. This has been discussed in Refs.~\cite{witten,tom}. In order 
to match phenomenologically preferred values for $\alpha_{\rm GUT}$, $M_{\rm GUT}$ and $M_{\rm p}$,
one would require the six dimensional volume of the Calabi-Yau manifold to be 
$V \sim (3\times 10^{16}{\rm GeV})^{-6}$ and the eleventh dimensional segment to be around 
$\pi r\sim (4\times 10^{15}{\rm GeV})^{-1}$. This automatically fixes the five dimensional Planck mass to
$M_{5}\sim 10^{17}$ GeV. In fact it turns out that following the limit posed in Eq.~(\ref{cond}),
the brane tension would dominate the energy density at late times, and the Universe would evolve 
like in the standard case. For such a large value of $M_{5}$ chaotic inflation could also be 
problematic, because the inflaton field would eventually take a value more than the four dimensional
Planck mass and thus the non-renormalizable terms could occur and spoil the inflationary potential.
If the reheating happens at a low energy scale then it is possible to escape the gravitino abundance 
bounds as set in Eq.~(\ref{life1}). It is obvious then that the transition from non-conventional 
to the standard cosmology happens much before the Universe thermalizes. Some of the interesting 
cosmological aspects have been discussed
in this regard\cite{lukas1,choi,tiago}. Especially in Ref.~\cite{tiago}, the authors have considered the 
moduli problem in the context of hetrotic M-theory. Usually moduli have large vacuum expectation
values and mass of the order of the gravitino mass $m_{3/2}$. Cosmological problems concerning
light weakly interacting particles have been  discussed earlier in the context 
of supergravity Ref.~\cite{cou} and superstrings \cite{nano,carlos}. It has been recognized that
similar to a gravitino problem, the moduli should also decay before nucleosynthesis, which provides
the lower bound on their masses, and if they do not decay then they have an upper bound on their 
masses to avoid closing the Universe. It is worth mentioning here that the cosmological bounds 
on the mass of the moduli will remain be the same as in the standard Big Bang cosmology.

{\it Case $2 \rightarrow$} $M_{11} \geq 10^{7}$ GeV. This is the case when there is a upper bound 
on $r$ from the experiments on gravitational forces beyond $1$mm size \cite{caceres,antoniadis}.
In this case $10^{16}{\rm GeV}\leq M_{5} \leq 10^{7}{\rm GeV}$. In this case we can certainly
expect the deviation from the standard cosmology. The transition from non-conventional
to the standard cosmology can be estimated to be happening at a temperature $10^{15}{\rm GeV}
\leq {\rm T_{\rm transit}} \leq 10 {\rm GeV}$. The gravitino abundance after the end of inflation
would be a cause of major problem for $10^{14}{\rm GeV}\leq M_{5}$. Moduli 
and dilaton problems would be revisited in this regime, but the problems associated to them  
can be ameliorated if their masses are above $1$ TeV. However, their dynamics will be governed 
by the non-conventional cosmology. 

{\it Case $3 \rightarrow$} $M_{11} \geq 1$TeV with $1/r \ll 1$TeV, this case has been discussed in
Refs.~\cite{benakli,thomas}. For $M_{11} \sim 1$ TeV, and $V \sim 1.7$ GeV \cite{benakli2}, 
the value of the five dimensional Planck mass is $M_{5} \sim 10^{9}$ GeV. This leads to the 
transition temperature $T_{\rm transit}\sim 10^{4}$ GeV. Thus this would also lead to non-conventional
cosmology and the gravitino abundance would cause a problem for nucleosynthesis provided
$M_{5} \geq 10^{14}$ GeV. However, for smaller values of $M_{5}$ the problem could be evaded.


\section{Discussions}

In this paper we have discussed some consequences of brane cosmology. Our inference is 
based upon the fact that the Friedmann equation in the early and the late cosmology 
could deviate from the standard one due to the presence of an extra dimension compactified 
on an orbifold. We have given more emphasis upon post-inflationary cosmology.
We have assumed that we had a period of inflation governed by the quadratic potential
whose mass is constrained by the COBE normalization. We also assume that supersymmetry is 
required to solve the hierarchy problem. This is quite natural if the string scale is more than
TeV. In this regard we have noticed two important points. 
It is not always true that post-inflationary brane cosmology will deviate from the standard
cosmology. If the decay rate of the inflaton is very slow, especially if they decay via four dimensional
Planck mass suppressed couplings, then no matter whatsoever be the five dimensional Planck mass is,
the Universe will end up with radiation energy density much less than the brane tension.
This would automatically lead to the standard cosmology. In the opposite limit, if the inflaton
decays fast enough via Yukawa or gauge couplings then the gravitinos produced from the thermal 
bath would have abundance crucially depending upon the five dimensional Planck mass. As we know 
that gravitinos decay very late, thus their decay products could generate enough entropy to 
wash out previously obtained baryon asymmetry and also harming synthesis of light elements. 
This is the reason why the gravitino abundance is a cause of great concern in any cosmological set-up. 
In this paper we have put some constraints upon the five dimensional Planck mass from the 
gravitino constraints coming from nucleosynthesis. We have noticed that the Universe undergoes 
a transition from non-conventional cosmological evolution to the standard evolution
before nucleosynthesis takes place, and it happens at a temperature which again depends 
upon the five dimensional Planck mass. Interestingly this transition could happen very 
late provided the five dimensional Planck mass is small enough. If it happens after  
supersymmetry breaking scale in the observable
world then there could be some interesting consequences, such as extremely low abundance of gravitinos,
and, realization of Affleck-Dine baryogenesis at very low temperatures. There could be many other
interesting scenarios which we have not discussed here such as entropy generation from
the decay of moduli could be ameliorated. Finally we studied the Affleck-Dine baryogenesis in some 
detail. We have noticed that such a scheme could be realizable only if the inflaton decays very slowly.
If this is so then while the inflaton is oscillating and before it decays the Universe undergoes
a transition from non-conventional to the standard one. This makes the overall discussion
bit more complicated, but nonetheless it is possible to realize such a baryogenesis.

\section{ACKNOWLEDGEMENTS}

The author is supported by the INLAKS foundation. The author is grateful to Andrew Liddle and
Andr\'e Lukas for helpful discussions.


\end{document}